\documentclass{SVProc}
\usepackage{makeidx}  
\makeindex
\usepackage{url}
\usepackage{xspace}
\usepackage[svgnames]{xcolor}
\usepackage{subfiles}
\usepackage{booktabs}
\usepackage{subfiles}
\usepackage{graphicx}
\usepackage{multicol}
\usepackage{subcaption}
\usepackage{array}
\usepackage{url}
\usepackage{mathtools}
\usepackage{chngcntr}
\usepackage{float}
\usepackage{anyfontsize}
\usepackage{cuted,tcolorbox,lipsum}
\usepackage{listings}
\usepackage{framed}
\usepackage{mciteplus}

\usepackage{array, booktabs, makecell}
\usepackage{siunitx, mhchem}
\usepackage{csquotes}

\usepackage{framed}
\usepackage{mdframed}

\definecolor{mypink2}{rgb}{1,1,1}

\usepackage{bm,amsmath}
\usepackage{amsfonts} 
\usepackage{physics}
\usepackage{tabularx}
\usepackage{array}

\newcolumntype{Y}{>{\centering\arraybackslash}X}

\usepackage{blindtext}

\usepackage{hyperref} %

\begin{document}
\frontmatter          
\pagestyle{headings}  
%

%
\mainmatter              
\title{An Assessment of ChatGPT on Log Data}
\titlerunning{An Assessment of ChatGPT on Log Data}  
%
\author{Priyanka Mudgal\inst{*} \and Rita Wouhaybi\inst{*}}
\authorrunning{Mudgal et al.} 
%
\tocauthor{Priyanka Mudgal and Rita Wouhaybi}
\index{Mudgal, P.}
\index{Wouhaybi, R.}

\institute{Intel Corporation, Hillsboro OR 97124, USA,\\
\email{*priyanka.mudgal@intel.com, \\ *rita.h.wouhaybi@intel.com}
}

\maketitle              

\begin{abstract}
Recent development of large language models (LLMs), such as ChatGPT has been widely applied to a wide range of software engineering tasks. Many papers have reported their analysis on the potential advantages and limitations of ChatGPT for writing code, summarization, text generation, etc. However, the analysis of the current state of ChatGPT for log processing has received little attention. Logs generated by large-scale software systems are complex and hard to understand. Despite their complexity, they provide crucial information for subject matter experts to understand the system status and diagnose problems of the systems. In this paper, we investigate the current capabilities of ChatGPT to perform several interesting tasks on log data, while also trying to identify its main shortcomings. Our findings show that the performance of the current version of ChatGPT for log processing is limited, with a lack of consistency in responses and scalability issues. We also outline our views on how we perceive the role of LLMs in the log processing discipline and possible next steps to improve the current capabilities of ChatGPT and the future LLMs in this area. We believe our work can contribute to future academic research to address the identified issues.
\end{abstract}
\section{Introduction}
In recent years, the emergence of generative AI and large language models (LLMs) such as OpenAI's ChatGPT have led to significant advancements in NLP. Many of these models provide the ability to be fine-tuned on custom datasets \cite{wang-etal-2018-glue}, \cite{mesh-transformer-jax}, \cite{gpt-j} and achieve the state-of-the-art (SOTA) performance across various tasks. A few of the LLMs such as GPT-3 \cite{brown2020language} have demonstrated in-context-learning capability without requiring any fine-tuning on task-specific data. The impressive performance of ChatGPT and other LLMs \cite{workshop2023bloom, tay2023ul2, thoppilan2022lamda, fedus2022switch, hoffmann2022training, zeng2022glm130b} in zero-shot and few-shot learning scenarios is a major finding as this helps LLMs to be more efficient \cite{855856, Lampert2009LearningTD, 10.5555/2984093.2984252, 10.5555/1620163.1620172, 10.5555/1620163.1620201}. With such learning methodologies, the LLMs can be used as a service \cite{sun2022blackbox} to empower a set of new real-world applications.

Despite the impressive capability of ChatGPT in performing a wide range of challenging tasks, there remain some major concerns about it in solving real-world problems like log analysis \cite{chatgpt_blog}. Log analysis is a vast area, and much research has been done. It mainly comprises three major categories, namely, log parsing, log analytics, and log summarization. Log parsing is an important initial step of system diagnostic tasks. Through log parsing, the raw log messages are converted into a structured format while extracting the template \cite{10.1145/3510003.3510101, zhu2019tools, 8029742, Du2016SpellSP}. Log analytics can be used to identify the system events and dynamic runtime information, which can help the subject matter experts to understand system behavior and perform system diagnostic tasks, such as anomaly detection \cite{10.1007/s10515-022-00370-w, 10.1109/ASE51524.2021.9678773, 10.1145/3338906.3338931, 10.1145/3133956.3134015}, log classification \cite{RAMACHANDRAN20231722}, error prediction \cite{Russo2015MiningSL, 10.1145/3208040.3208051}, and root cause analysis \cite{7794329, 8029786}. Log analytics can further be used to perform advanced operations e.g., identify user activities, and security analysis e.g., detect logged-in users, API/service calls, malicious URLs, etc. As logs are huge in volume, log summarization enables the operators to provide a gist of the overall activities in logs and empowers the subject matter experts to read and/or understand logs faster. Recent studies leverage pre-trained language models \cite{10.1109/ASE51524.2021.9678773, tao2022logstamp, le2023log} for representing log data. However, these methods still require either training the models from scratch \cite{Liu_2022} or tuning a pre-trained language model with labeled data \cite{10.1109/ASE51524.2021.9678773, le2023log}, which could be impractical due to the lack of computing resources and labeled data.

\begin{figure}
\centering
\includegraphics[width=0.5\textwidth]{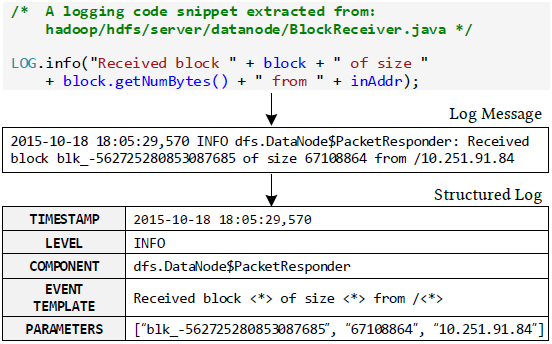}
\caption{\label{fig:log_example}An example of log code, log message, and structured log from \cite{he2020loghub}}
\end{figure}

More recently, LLMs such as ChatGPT \cite{chatgpt_blog} have been applied to a variety of software engineering tasks and achieved satisfactory performance \cite{10196869, cao2023study}. With a lack of studies to analyze ChatGPT's capabilities on log processing, it is unclear whether it can be performed well on the logs. Although many papers have performed the evaluation of ChatGPT on software engineering tasks \cite{laskar2023systematic, 10.1007/s10270-023-01105-5, sridhara2023chatgpt}, specific research is required to investigate its capabilities in system log area. We are aware that the LLMs are fast evolving, with new models, versions, and tools being released frequently, and each one is improved over the previous ones. However, our goal is to assess the current situation and to provide a set of experiments that can enable the researchers to identify possible shortcomings of the current version for analyzing logs and provide a variety of specific tasks to measure the improvement of future versions. Hence, in this paper, we conduct an initial level of evaluation of ChatGPT on log data. Specifically, we divide the log processing \cite{he2022empirical} into three subsections: log parsing, log analytics, and log summarization. We design appropriate prompts for each of these tasks and analyze ChatGPT's capabilities in these areas. Our analysis shows that ChatGPT achieves promising results in some areas, but limited outcomes in others and contains several real-world challenges in terms of scalability. In summary, the major contributions of our work are as follows:

• To the best of our knowledge, we are the first to study and analyze ChatGPT’s ability to analyze the log data in multiple detailed aspects.

• We design the prompts for multiple scenarios in log processing and record ChatGPT's response.

• Based on the findings, we outline several challenges and prospects for ChatGPT-based log processing.

\vspace{-1.0 cm}
\begin{figure*}

 
  
\begin{subfigure}{.475\linewidth}
 {{\includegraphics[width=\linewidth, trim = 0cm 10cm 0cm 0cm]{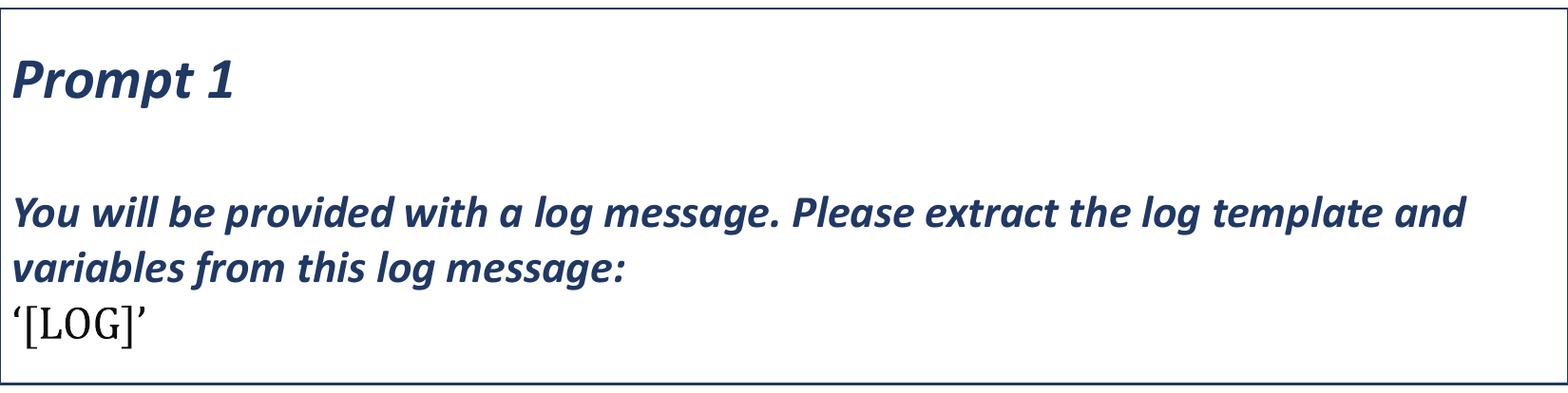} }}%
 
  \label{velcomp}
\end{subfigure}\hfill 
\begin{subfigure}{.475\linewidth}
  {{\includegraphics[width=\linewidth, trim = 0cm 10cm 0cm 0cm]{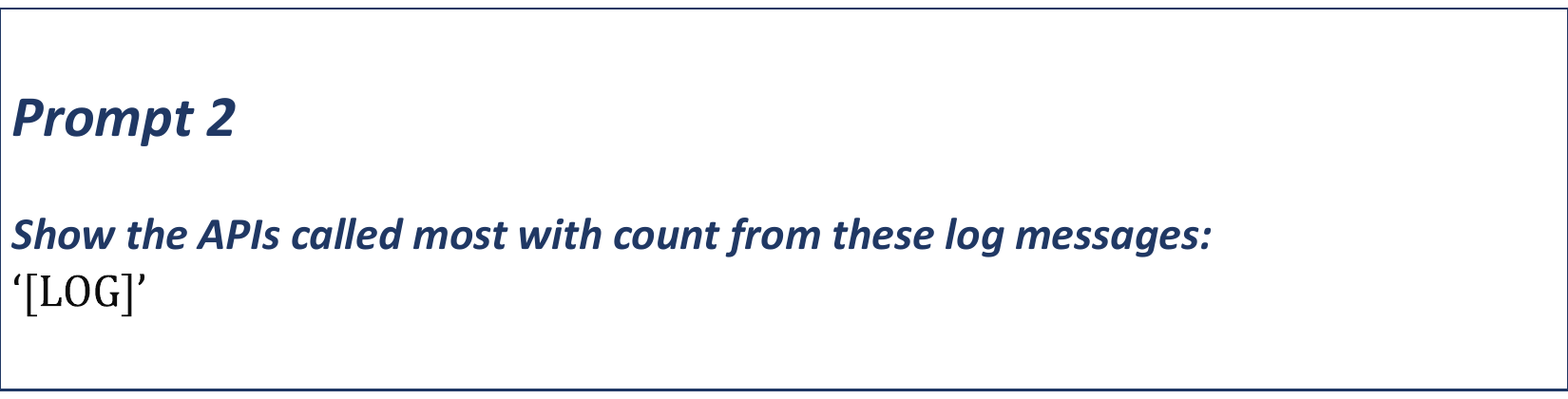} }}%
  
  \label{estcomp}
\end{subfigure}

\medskip 
\vspace{-1.0 cm}
\begin{subfigure}{.475\linewidth}
 {{\includegraphics[width=\linewidth, trim = 0cm 10cm 0cm 0cm]{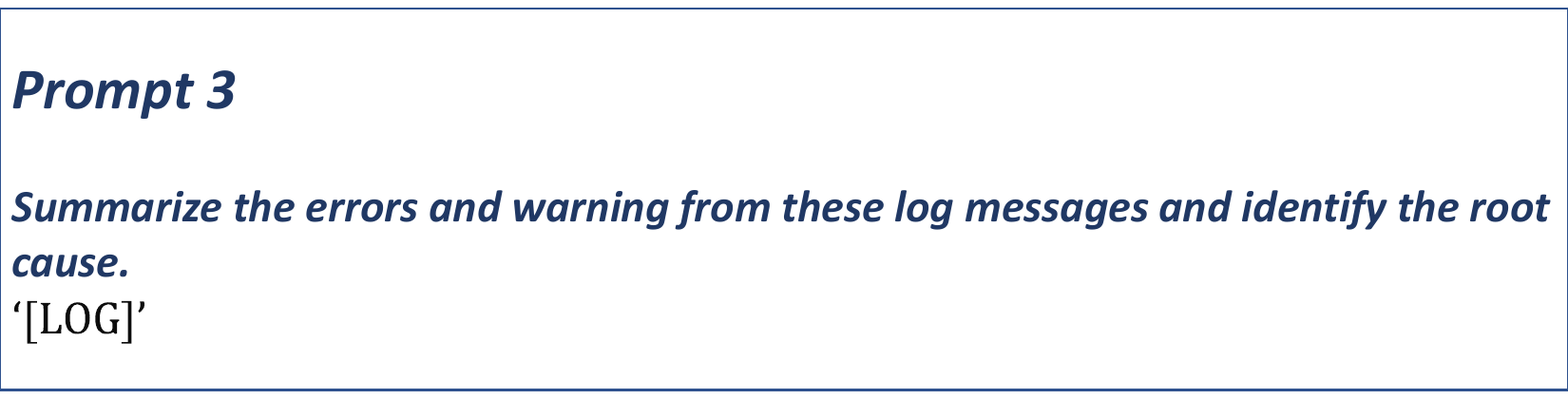} }}%
 
  \label{velcomp}
\end{subfigure}\hfill 
\begin{subfigure}{.475\linewidth}
  {{\includegraphics[width=\linewidth, trim = 0cm 10cm 0cm 0cm]{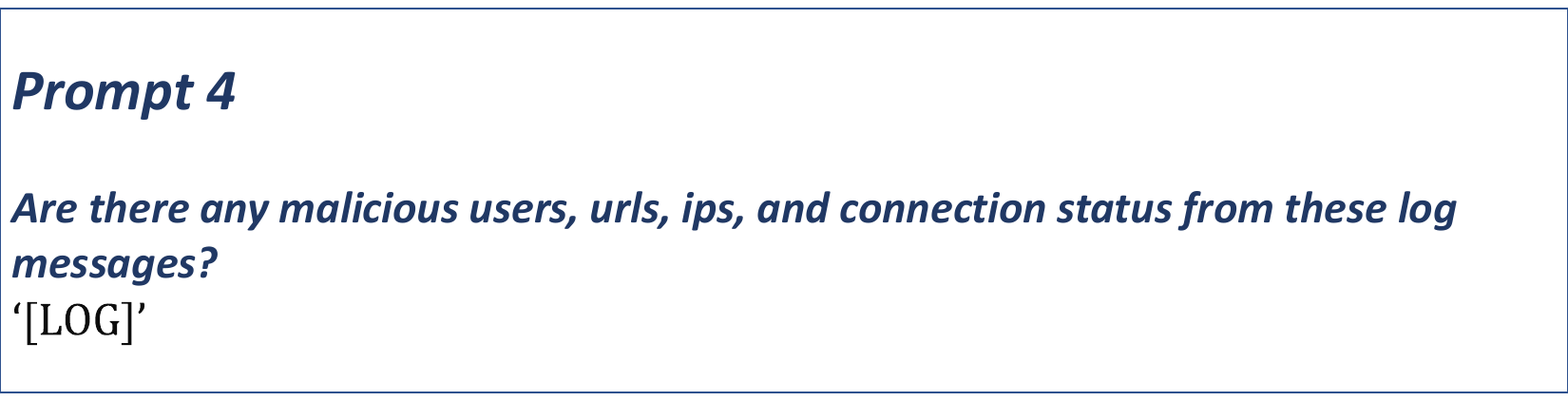} }}%
  
  \label{estcomp}
\end{subfigure}

\medskip 
\vspace{-1.0 cm}
\begin{subfigure}{.475\linewidth}
 {{\includegraphics[width=\linewidth, trim = 0cm 10cm 0cm 0cm]{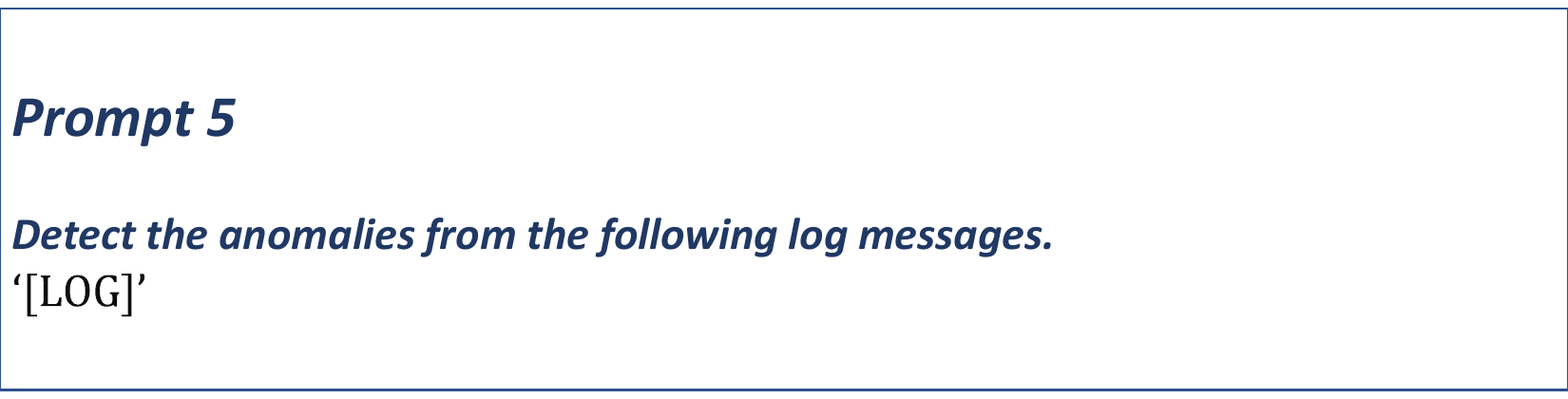} }}%
 
  \label{velcomp}
\end{subfigure}\hfill 
\begin{subfigure}{.475\linewidth}
  {{\includegraphics[width=\linewidth, trim = 0cm 10cm 0cm 0cm]{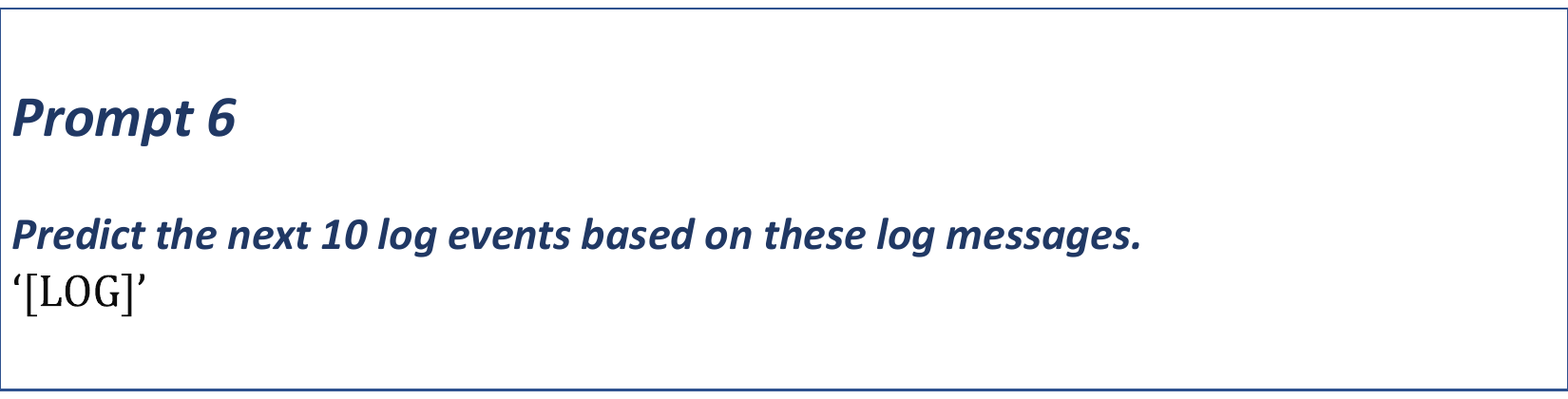} }}%
  
  \label{estcomp}
\end{subfigure}

\medskip 
\vspace{-1.0 cm}
\begin{subfigure}{.475\linewidth}
 {{\includegraphics[width=\linewidth, trim = 0cm 10cm 0cm 0cm]{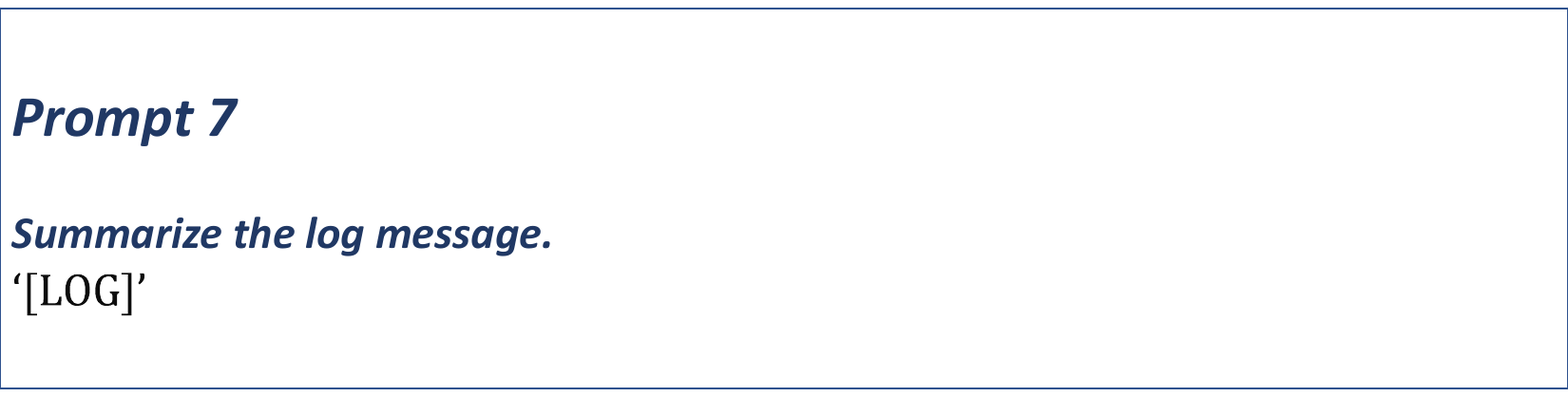} }}%
 
  \label{velcomp}
\end{subfigure}\hfill 

\caption{Various prompt designs to address the research questions.}
\label{fig:prompts}
\end{figure*}

\vspace{-1.0 cm}
\section{\textbf{Related Work}}
\subsection{\textbf{Log data}}
With the increasing scale of software systems, it is complex to manage and maintain them. To tackle this challenge, engineers enhance the system observability \cite{oreilly_book, observability} with logs. 

Logs capture multiple system run-time information such as events, transactions, and messages. A typical piece of log message is a time-stamped record that captures the activity that happened over time (e.g., software update events or received messages). Logs are usually generated when a system executes the corresponding logging code snippets. An example of the code snippet and generated code is shown in Fig. \ref{fig:log_example}. A system with mature logs essentially facilitates the system behavior understanding, health monitoring, failure diagnosis, etc. Generally, there are three standard log formats, i.e., structured, semi-structured, and unstructured logs \cite{GANDOMI2015137}. These formats share the same components: a timestamp and a payload content.

Structured logs usually keep a consistent format within the log data and are easy to manage. Specifically, the well-structured format allows easy storing, indexing, searching, and aggregation in a relational database. The unstructured log data achieves its high flexibility at the expense of the ease of machine processing. The characteristic of free-form text becomes a major obstacle for efficient query and analysis on unstructured or semi-structured logs. For instance, to count how often an API version appears in unstructured logs, engineers need to design a complex query with ad-hoc regular expressions to extract the desired information. The manual process takes lots of time and effort and is not scalable.

\subsection{\textbf{Log Processing}}
Logs have been widely adopted in software system development and maintenance. In industry, it is a common practice to record detailed software runtime information into logs, allowing developers and support engineers to track system behaviors and perform postmortem analysis. On a high level, log processing can be categorized in three types as discussed below.
\subsubsection{\textbf{Log Parsing}}
Log parsing is generally the first step toward automated log analytics. It aims at parsing each log message into a specific log event/template and extracting the corresponding parameters. Although there are many traditional regular expression-based log parsers, but, they require a predefined knowledge about the log template. To achieve better performance in comparison to traditional log parsers, many data-driven \cite{1251233, 10.1145/3540250.3549176, 5463281, dai2020logram, Shima2016LengthMC, 10.1145/2063576.2063690, 8029742} and deep learning based approaches \cite{le2023log, Liu_2022} have been proposed to automatically distinguish template and parameter parts.
\subsubsection{\textbf{Log Analytics}} \label{log_Analytics_related_work}
Modern software development and operations rely on log monitoring to understand how systems behave in production. There is an increasing trend to adopt artificial intelligence to automate operations. Gartner \cite{gartner_2019} refers to this movement as AIOps.  The research community, including practitioners, has been actively working to address the challenges related to extracting insights from log data also being referred to as “Log Analysis” \cite{log_summarization}. Various insights that can be gained are in terms of log mining \cite{PETTINATO2019121}, error detection and root cause analysis, security and privacy, anomaly detection, and event prediction.

\par{\textbf{Log Mining}} Log mining seeks to support understanding and analysis utilizing abstraction and extracting useful insights. However, building such models is a challenging and expensive task. In our study, we confine ourselves to posing specific questions in terms of most API/service calls that can be extracted out of raw log messages. This area is well studied from a deep learning aspect and most of those approaches \cite{steinle2006mapping, 4700316, awad2016performance, 1174915, lou2010mining, tan2010visual, beschastnikh2014inferring, kc2011elt} require to first parse the logs and then process them to extract the detailed level of knowledge.
\par{\textbf{Error Detection and Root Cause Analysis}} Automatic error detection from logs is an important part of monitoring solutions. Maintainers need to investigate what caused that unexpected behavior. Several studies \cite{kimura2014spatio, ren2019root, pi2018profiling, chuah2013linking, zheng2011co, gurumdimma2016crude} attempt to provide their useful contribution to root cause analysis, accurate error identification, and impact analysis.

\par{\textbf{Security and Privacy}} Logs can be leveraged for security purposes, such as malicious behaviour and attack detection, URLs, and IP detection, logged-in user detection, etc. Several researchers have worked towards detecting early-stage malware and advanced persistence threat infections to identify malicious activities based on log data \cite{oprea2015detection, gonccalves2015big, balzarotti2012research, yoon2014toward, yen2013beehive}.

\par{\textbf{Anomaly Detection}} Anomaly detection techniques addresses to identify the anomalous or undesired patterns in logs. The manual analysis of logs is time-consuming, error-prone, and unfeasible in many cases. Researchers have been trying several different techniques for automated anomaly detection, such as deep learning \cite{zhang2019robust, du2017deeplog, bertero2017experience, meng2019loganomaly} and data mining, statistical learning methods, and machine learning \cite{xu2009online, xu2009detecting, lim2008log, tang1992analysis, candidolog, he2016experience, lu2017log}.

\par{\textbf{Event Prediction}} The knowledge about the correlation of multiple events, when combined to predict the critical or interesting event is useful in preventive maintenance or predictive analytics that can reduce the unexpected system downtime and result in cost saving \cite{189e0e3f-95de-38a8-b475-180e6690e629, 6424841, TAMA2019233}. Thus, the event prediction method is highly valuable in real-time applications. In recent years, many rule-based and deep learning based approaches \cite{10.1007/978-3-319-59536-8_30, evermann2017predicting, verenich2018survey, Francescomarino2017AnEI, maggi2013predictive, 10.1007/978-3-642-13094-6_5} have evolved and performing significantly.

\subsubsection{\textbf{Log Summarization}}
Log statements are inserted in the source code to capture normal and abnormal behaviors. However, with the growing volume of logs, it becomes a time-consuming task to summarize the logs. There are multiple deep learning-based approaches \cite{meng2020summarizing, log_summarization, log_summarization_zebrium, RAMACHANDRAN20231722} that perform the summarization, but they require time and compute resources for training the models.

\subsection{\textbf{ChatGPT}}
ChatGPT is a large language model which is developed by OpenAI \cite{chatgpt_blog, chatgpt_ui}. ChatGPT is trained on a huge dataset containing massive amount of internet text. It offeres the capability to generate text responses in natural language that are based on a wide range of topics. The fundamental of ChatGPT is generative pre-training transformer (GPT) architecture. GPT architecture is highly effective for natural language processing tasks such as translation in multiple languages, summarization, and question answering (Q \& A). It offers the capability to be fine-tuned on specific tasks with a smaller dataset with specific examples. ChatGPT can be adopted in a variety of use cases including chatbots, language translation, and language understanding. It is a powerful tool and possesses the potential to be used across wide range of industries and applications.
\subsection{\textbf{ChatGPT Evaluation}}
Several recent works on ChatGPT evaluation have been done, but most of the papers target the evaluations on general tasks \cite{chang2023survey, sridhara2023chatgpt}, code generation \cite{10196869}, deep learning-based program repair \cite{cao2023study}, benchmark datasets from various domains \cite{laskar2023systematic}, software modeling tasks \cite{10.1007/s10270-023-01105-5}, information extraction \cite{li2023evaluating}, sentiment analysis of social media and research papers \cite{leiter2023chatgpt} or even assessment of evaluation methods \cite{mao2023gpteval}. The closest to our work is \cite{le2023evaluation}, but they focus only on log parsing.

\par{We believe that the log processing area is huge and a large-level evaluation of ChatGPT on log data would be useful for the research community. Hence, in our work, we focus on evaluating ChatGPT by conducting an in-depth and wider analysis of log data in terms of log parsing, log analytics, and log summarization.}

\section{\textbf{Context}}

In this paper, our primary focus is to assess the capability of ChatGPT on log data. In line with this, we aim to answer several research questions through experimental evaluation.

\subsection{\textbf{Research Questions}}

\subsubsection{\textbf{Log Parsing}}
\par{\textbf{RQ1.} How does ChatGPT perform on log parsing?}

\subsubsection{\textbf{Log Analytics}}
\par{\textbf{RQ2.} Can ChatGPT extract the errors and identify the
root cause from raw log messages?}
\par{\textbf{RQ3.} How does ChatGPT perform on advanced analytics tasks e.g., most called APIs/services?}
\par{\textbf{RQ4.} Can ChatGPT be used to extract security information
from log messages?}
\par{\textbf{RQ5.} Is ChatGPT able to detect anomalies from log data?}

\par{\textbf{RQ6.} Can ChatGPT predict the next events based on previous log messages?}

\subsubsection{\textbf{Log Summarization}}

\par{\textbf{RQ7.} Can ChatGPT summarize a single raw log messages?}
\par{\textbf{RQ8.} Can ChatGPT summarize multiple log messages?}

\subsubsection{\textbf{General}}
\par{\textbf{RQ9.} Can ChatGPT process bulk log messages?}
\par{\textbf{RQ10.} What length of log messages can ChatGPT process at once?}

To examine the effectiveness of ChatGPT in answering the research questions, we design specific prompts as shown in Fig \ref{fig:prompts}. We append the log messages in each of the prompts (in place of the slot '[LOG]').

\subsection{\textbf{Dataset}}

To perform our experiments, we use the datasets provided from the Loghub benchmark  \cite{zhu2019tools, he2020loghub}. This benchmark covers log data from various systems, including,  windows and linux operating systems, distributed systems, mobile systems, server applications, and standalone software. Each system dataset contains 2,000 manually labeled and raw log messages.
\begin{figure}[h]
    \centering
    \vspace{2.5 cm}
\includegraphics[width=\linewidth, trim = 0cm 0 0cm 10cm]{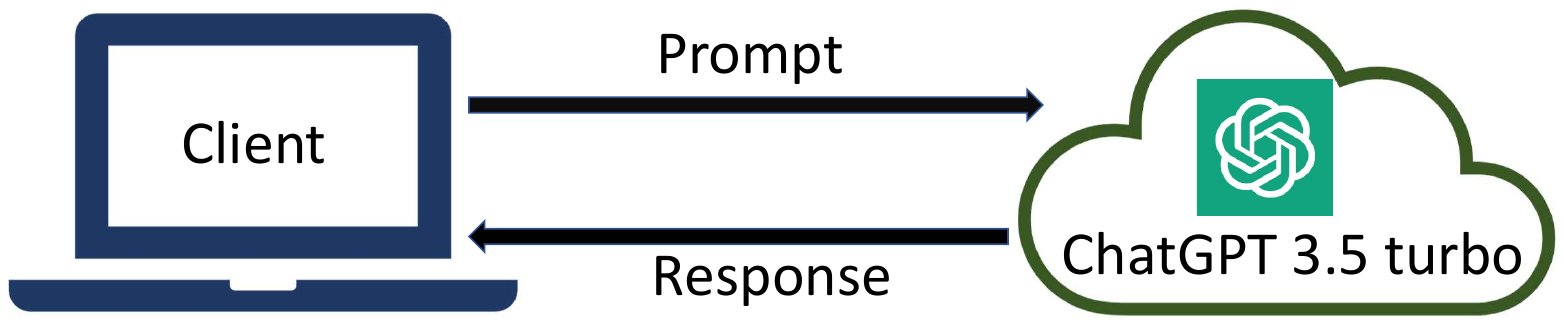}
    \label{fig:1}
\vspace{-5.5 cm}

    \caption{Flow Diagram.}
    \label{fig:flow}
\end{figure}
\vspace{-1.0 cm}
\subsection{\textbf{Experimental Setup}}
For our experiments, we are using the ChatGPT API based on the gpt-3.5-turbo model to generate the responses for different prompts \cite{chatgpt_blog}. As shown in Fig. \ref{fig:flow}, we send the prompts appended with log messages to ChatGPT from our system with Intel\textsuperscript{\textregistered} Xeon\textsuperscript{\textregistered} E3-1200 v5 processor and Intel\textsuperscript{\textregistered} Xeon\textsuperscript{\textregistered} E3-1500 v5 processor and receive the response. To avoid bias from model updates, we use a snapshot of gpt3.5-turbo from March 2023 \cite{chatgpt_3.5}.

\subsection{\textbf{Evaluation Metrics}}
As our study demands a detailed evaluation and in some cases, there was no state-of-the-art tool, we evaluated the output by our manual evaluation.

\section{\textbf{Experiments and Results}}
Each of the subsections below describes the individual evaluation of ChatGPT in different areas of log processing.

\begin{figure}[h]
    \centering
     \vspace{1.5 cm}
\includegraphics[width=\linewidth, trim = 0cm 16.2cm 0cm 10cm]{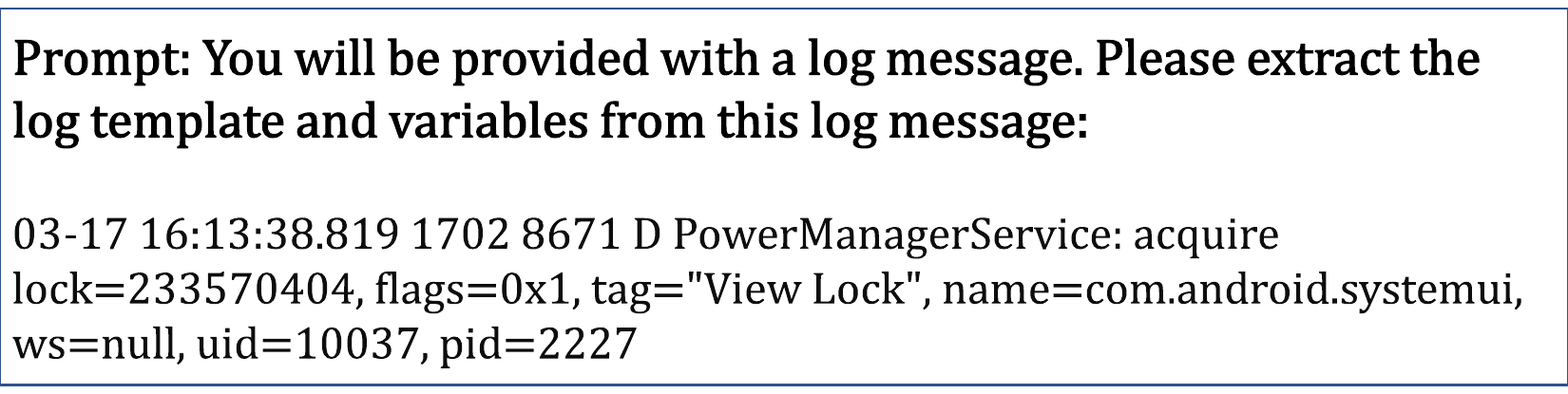}
    \label{fig:1}
\vspace{0.00mm} 
\includegraphics[width=\linewidth, trim = 0cm 5cm 0cm 0cm]{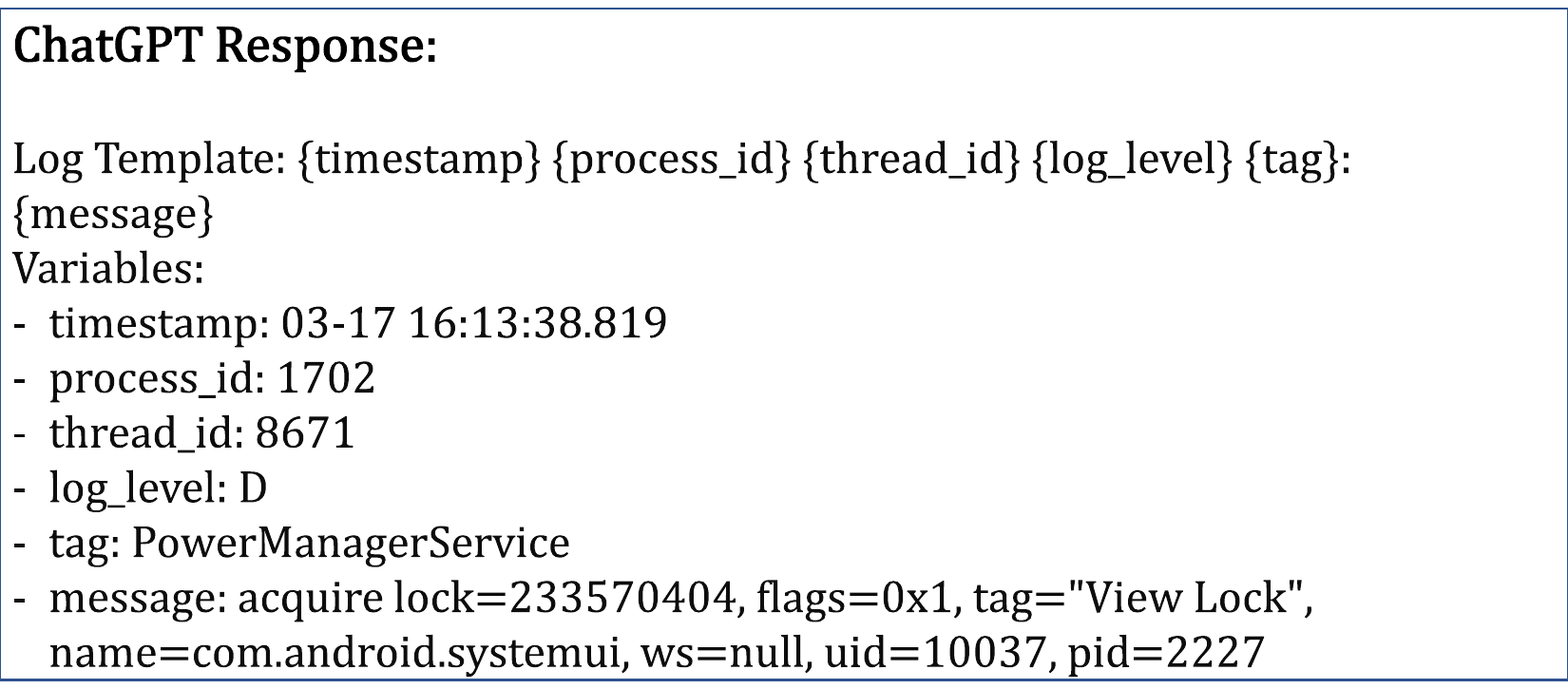}
\vspace{-1.0 cm}
    \caption{Log parsing of raw log message.}
    \label{fig:log_parsing_1example}
\end{figure}
\begin{figure}[h]
    \centering
    \vspace{1.0cm}%
\includegraphics[width=\linewidth, trim = 0cm 15.82cm 0cm 5cm]{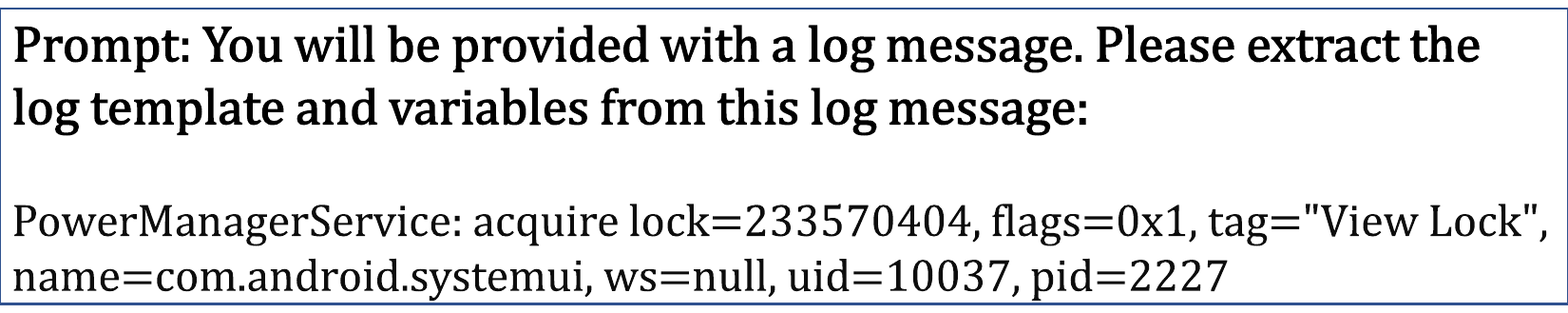}
    \label{fig:1}
\includegraphics[width=\linewidth, trim = 0cm 15cm 0cm 1.5cm]{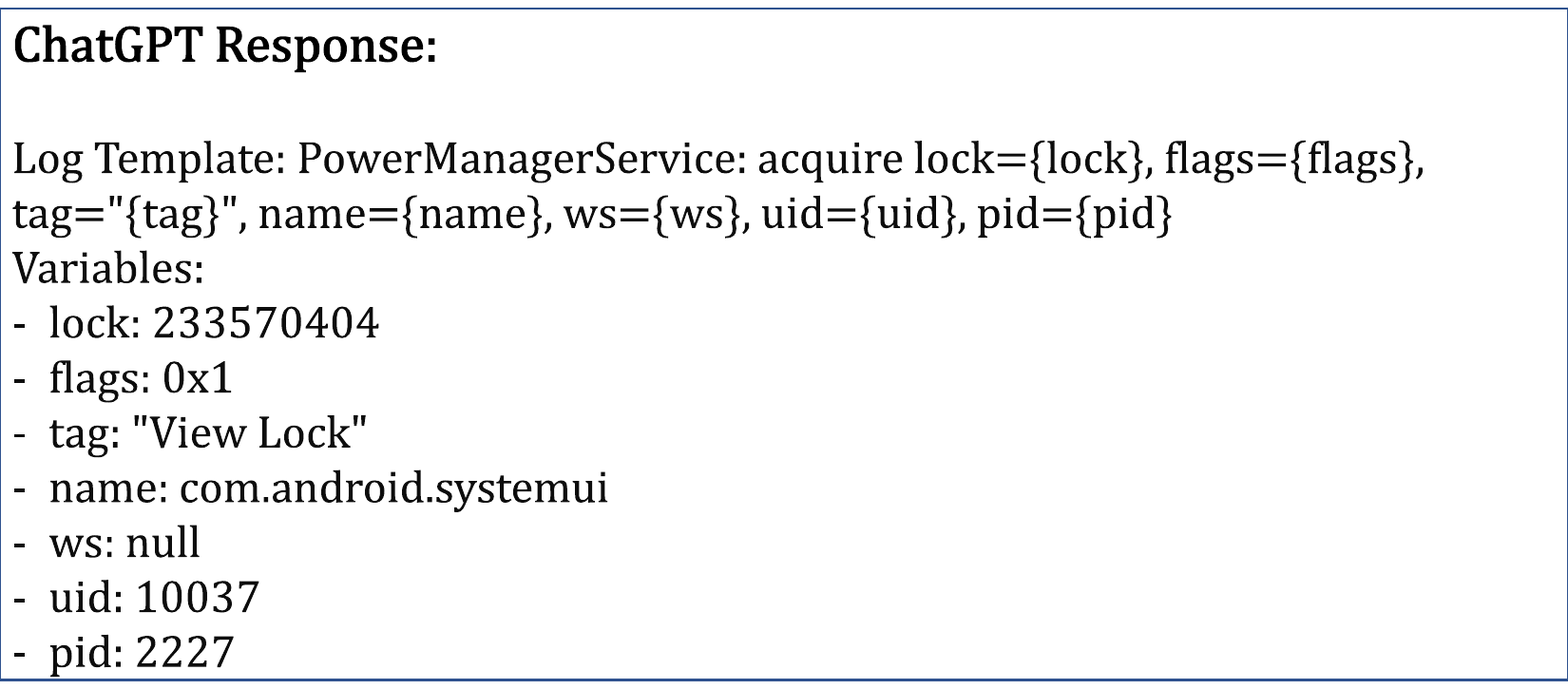}

    \vspace{4.0 cm}
    \caption{Log parsing of preprocessed log message.}
    \vspace{-1.0 cm}
    \label{fig:log_parsing_2example}
\end{figure}

\subsection{\textbf{Log Parsing}}
In this experiment, we assess the capability of ChatGPT in parsing a raw log message and a preprocessed log message and find the answer to \textbf{RQ1}. For the first experiment, we provide a single raw log message from each of the sixteen publicly available datasets \cite{he2020loghub} and ask ChatGPT to extract the log template. We refer to it as first-level log parsing. ChatGPT performs well in extracting the specific parts of log messages for all sixteen log messages. One of the examples of ChatGPT's response for first-level log parsing is shown in Fig. \ref{fig:log_parsing_1example}. Next, we preprocess the log message, extract the content, and ask chatGPT to further extract the template from the log message. ChatGPT can extract the template and variables from the log message successfully on all sixteen log messages with a simple prompt. One of the examples of ChatGPT's response is shown in Fig. \ref{fig:log_parsing_2example}.

\subsection{\textbf{Log Analytics}}
To evaluate ChatGPT's capability in log analytics, we perform several experiments in each of the categories described in section \ref{log_Analytics_related_work}.

\par
\begin{figure}[h]
    \centering
    \vspace{-15px}
\includegraphics[width=\linewidth, trim = 0cm 16.8cm 0cm 10cm]{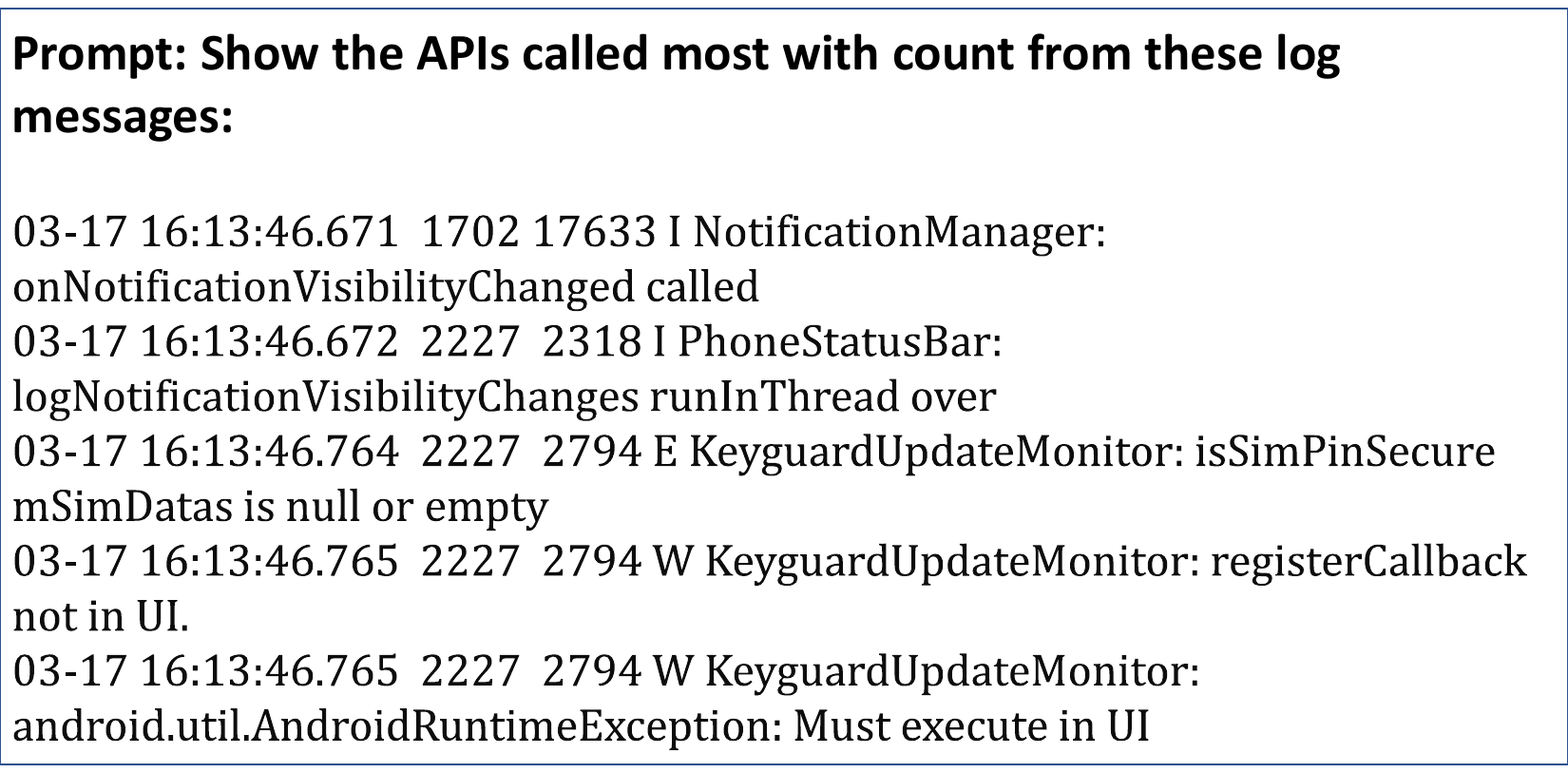}
    \label{fig:1}
 \vspace{3.0cm}
\includegraphics[width=\linewidth, trim = 0cm 18cm 0cm 2.15cm]{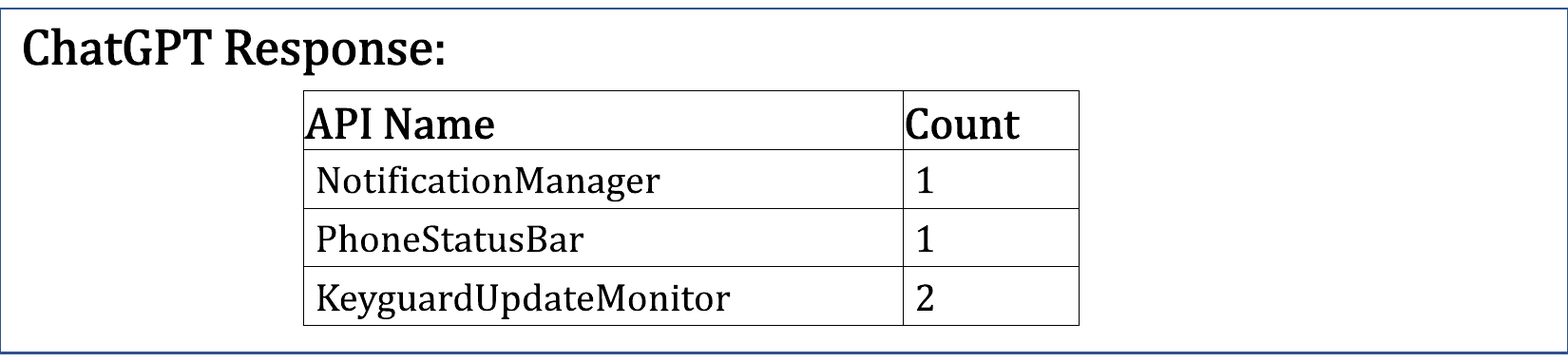}

    \vspace{2.5cm}%
    \caption{ChatGPT response to extract the APIs from log messages.}
    \label{fig:api_prompt}
\end{figure}

{\textbf{Log Mining}} In this experiment, we are seeking the answer of \textbf{RQ2} by investigating if ChatGPT can skim out the knowledge from raw logs without building an explicit parsing pipeline. We perform our experiments in several parts. We provide a subset of log messages containing 5, 10, 20, and 50 log messages from Loghub benchmark \cite{he2020loghub} and ask ChatGPT to identify the APIs. Fig \ref{fig:api_prompt} shows an example of ChatGPT response when a smaller set of log messages were passed. We notice that ChatGPT consistently missed identifying some APIs from the log messages irrespective of the count of log messages, but still shows 75\% or more accuracy in all cases. Results are reported in Table \ref{tab:api}.
 \vspace{-0.5 cm} 

\begin{table}[H]
\setlength{\tabcolsep}{1pt}
\caption{\textbf{ChatGPT's performance to identify the APIs, errors and root cause from Loghub dataset \cite{he2020loghub}.}}
\begin{center}
 \scriptsize
 \begin{tabular}{cccccccccc}\toprule

\thead{Log \\ Message \\ Count} & \thead{API \\ Count} & \thead{API \\captured} & \thead{API \\ Accuracy (\%)} & \thead{API \\ Response \\Time (s)} & \thead{Error \\ Count} & \thead{Error \\captured} & \thead{Error \\Accuracy\\ (\%)} & \thead{Error \\ Response \\time (s)} \\
 \midrule  \addlinespace
5 & 5 & 4 & 80 & 2.48 &  2 & 2 & 100 & 18.49\\ 
10 & 10 & 8 & 80 & 3.96 & 3 & 3 & 100 & 27.61 \\
20 & 20 & 15 & 75 & 6.44 & 5 & 3 & 60 & 36.38\\
50 & 50    & 46 & 92 & 5.66 & 13 & 5 & 38.46 & 46.46\\

 \bottomrule

\end{tabular}

\label{tab:api}
\end{center}
\end{table}

\begin{figure}[h]
    \centering
    \vspace{1.0cm}%
\includegraphics[width=\linewidth, trim = 0cm 15cm 0cm 5cm]{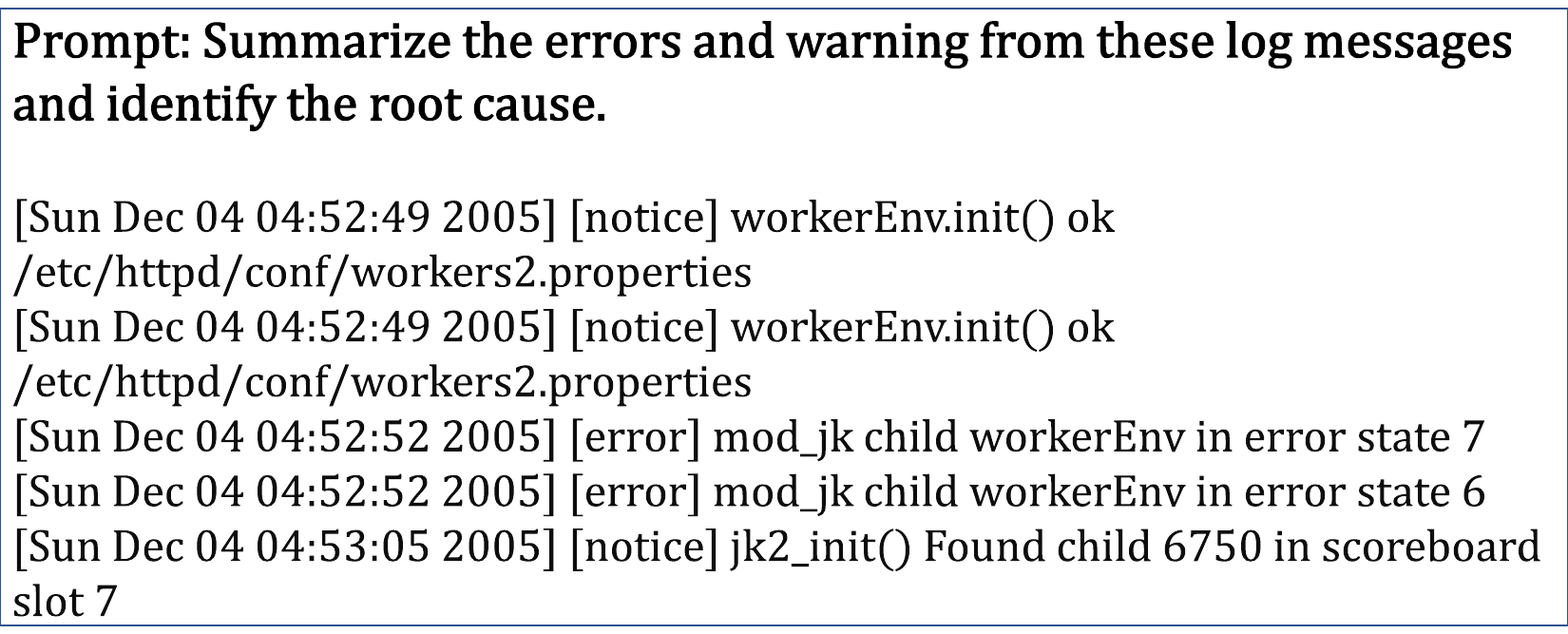}
    \label{fig:1}
    \vspace{2.55cm}
\includegraphics[width=\linewidth, trim = 0cm 5.5cm 0cm 5.2cm]{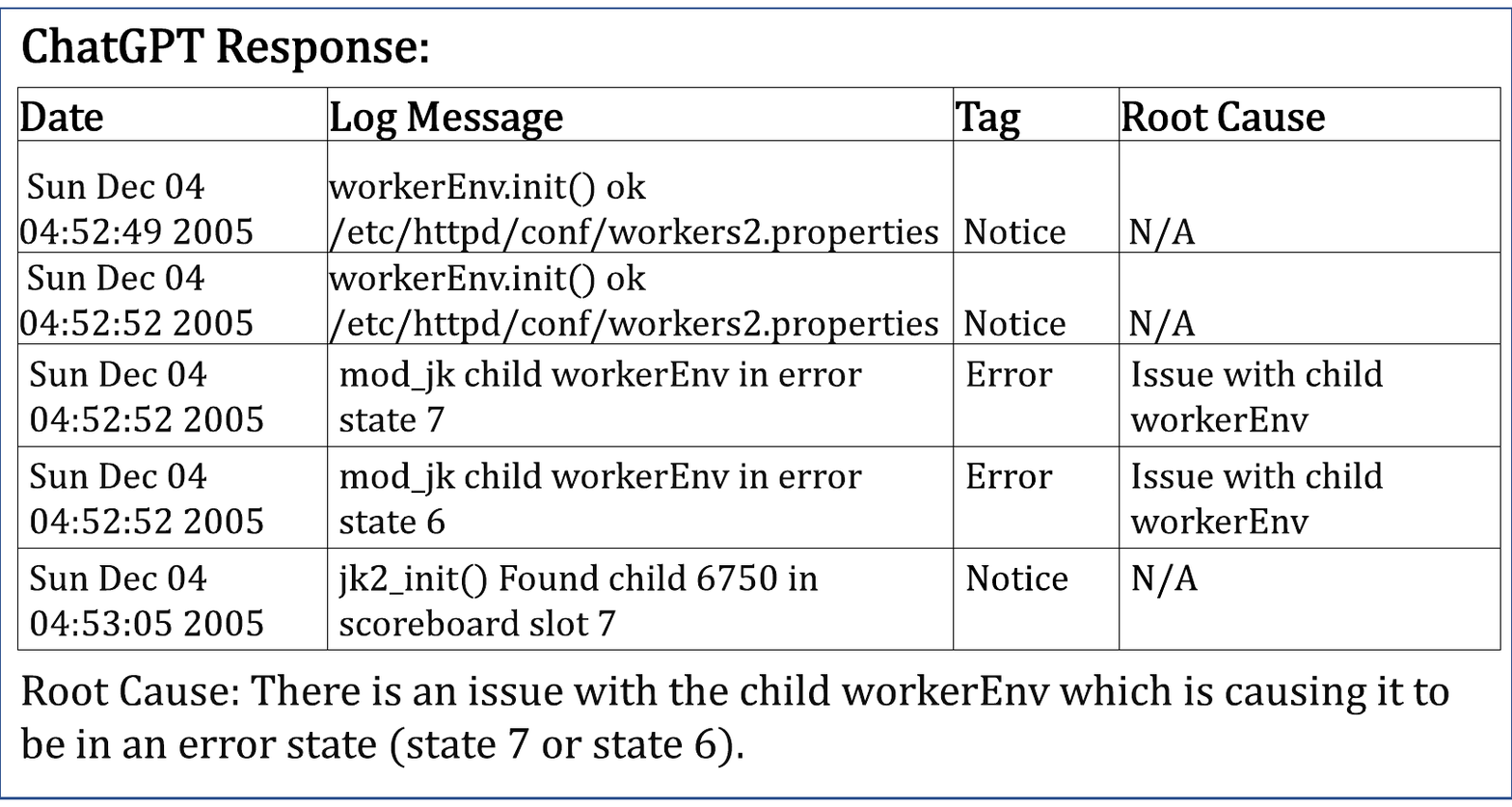}

    \vspace{20pt}%
    \caption{ChatGPT response to identify the errors and root cause from set of 5 log messages from Loghub dataset \cite{he2020loghub}.}
    \label{fig:error_example}
\end{figure}

\par{\textbf{Error Detection and Root Cause Analysis}}
In this experiment, we explicitly ask ChatGPT \cite{chatgpt_3.5} to identify the errors, warnings, and possible root causes of those in the provided log messages and address \textbf{RQ3}. Aligning towards our study structure, we first provide five log messages from the Loghub dataset \cite{he2020loghub} and later increase the size of log messages to ten, twenty, and fifty. Fig \ref{fig:error_example} shows the identified errors from five log messages and a detailed report for all the combinations with their response time is being reported in Table \ref{tab:api}. It is evident from Table \ref{tab:api} that ChatGPT successfully identifies the errors and warnings on a smaller set of log messages than a larger set.

\begin{figure}[h]
    \centering
    \vspace{20pt}%
\includegraphics[width=\linewidth, trim = 0cm 15cm 0cm 5cm]{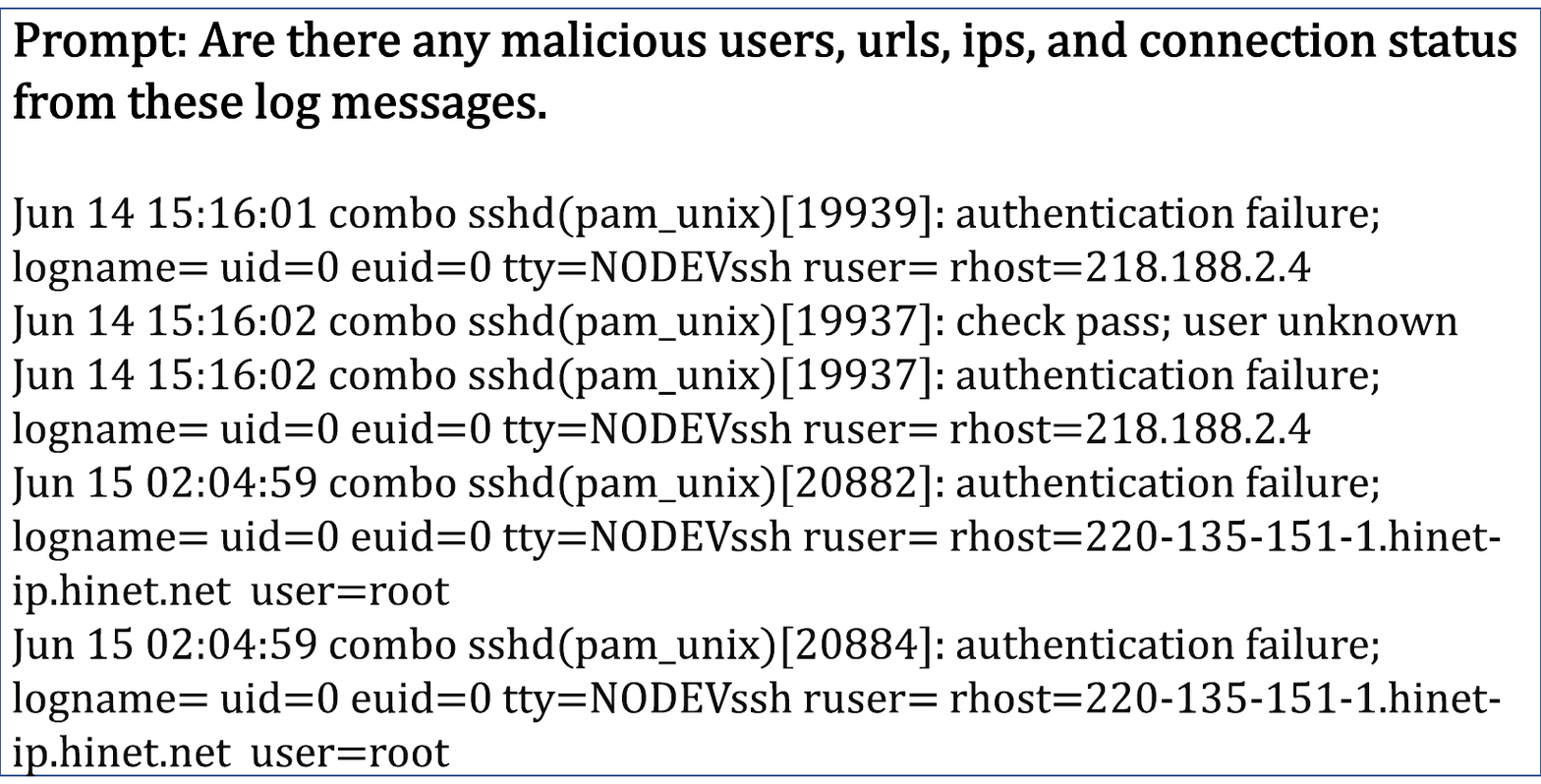}
    \label{fig:1}
 \vspace{2 cm}
\includegraphics[width=\linewidth, trim = 0cm 5cm 0cm 5cm]{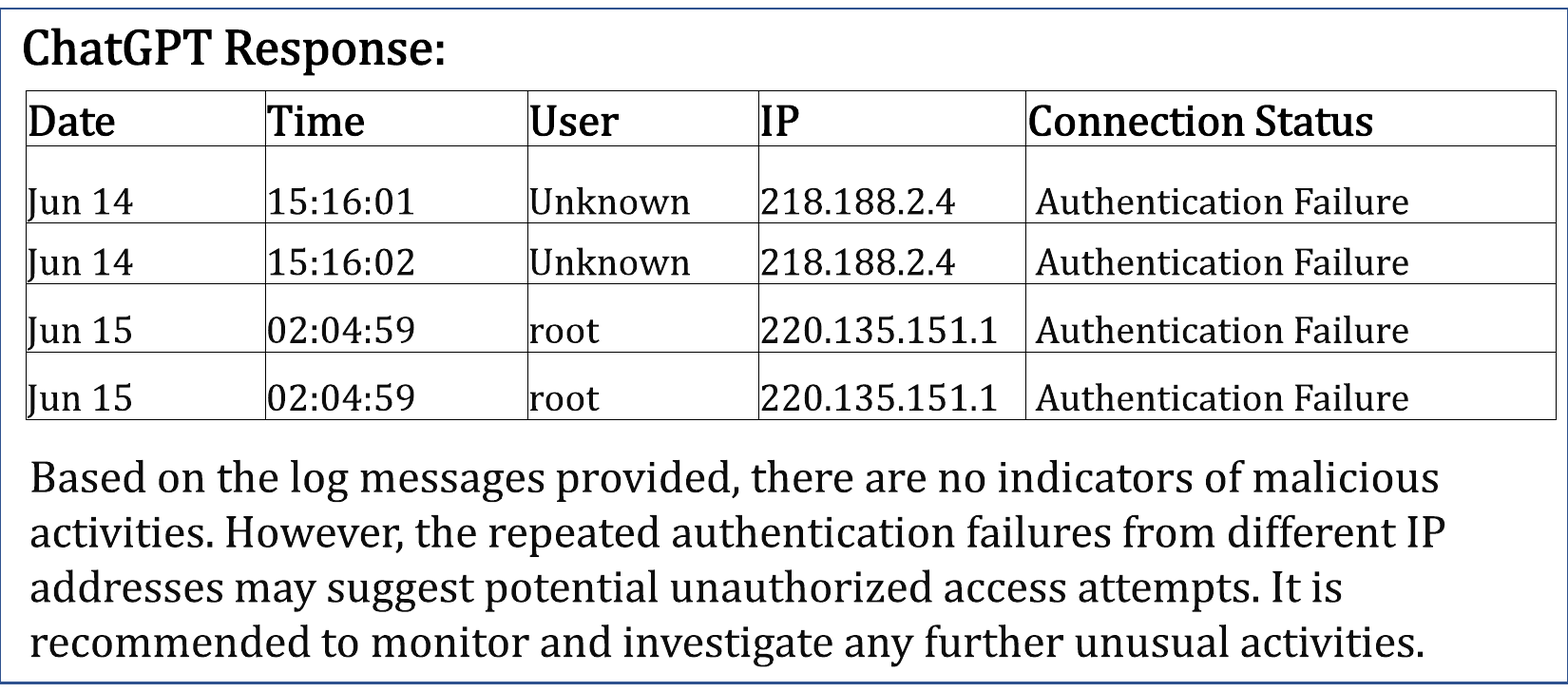}

    \vspace{-0.5cm}%
    \caption{ChatGPT response to extract urls, IPs, and users from set of 5 log messages from Loghub dataset \cite{Loghub}.}
    \label{fig:security_example}
\end{figure}

\par{\textbf{Security and privacy}} In this experiment, we focus on addressing \textbf{RQ4} and investigate if ChatGPT can identify the URLs, IPs, and logged users from the logs and extract knowledge about malicious activities. We use the open source dataset from Loghub \cite{Loghub} and follow the same approach of sending the set of five, ten, twenty, and fifty log messages to chatGPT to detect the URLs, IPs, and users from them. We use the 'Prompt 4' from Fig. \ref{fig:prompts} to ask if there are any malicious activities present in the logs. As shown in Table \ref{tab:security}, ChatGPT extracts out the IPs and logged-in users with high accuracy irrespective of the length of log messages. An example of ChatGPT's response is shown in Fig. \ref{fig:security_example}. The detailed report is shown in Table. \ref{tab:security}.

\vspace{-1.0cm}
\begin{table}[H]
\setlength{\tabcolsep}{1.7pt}
\caption{\textbf{ChatGPT performance to extract urls, IPs, and users from the log messages from Loghub dataset \cite{he2020loghub}.}}
\begin{center}
 \scriptsize
 \begin{tabular}{ccccccccc}\toprule
     
\thead{Log \\ Message \\ Count} & \thead{URLs \\ Count} & \thead{URLs \\captured}   & \thead{URL \\Accuracy \\ (\%)}   &  \thead{User \\ Count} & \thead{User \\captured} &  \thead{User \\ Accuracy \\ (\%)} & \thead{Response \\time (s)}\\
 \midrule  \addlinespace
5 & 4 & 4 & 100 & 2 &2  &100 & 13.77\\ 
10 & 9 & 9 & 100 & 7 &7  &100 & 46.41 \\
20 & 13 & 13 &100& 14 &14  &100 & 112.14\\
50 & 24 & 20& 83.33 & 16 &14  &87.5 & 163.76\\

 \bottomrule

\end{tabular}

\vspace{-1.0cm}
\label{tab:security}
\end{center}
\end{table}
\begin{figure}[h]
    \centering
    \vspace{20pt}%
\includegraphics[width=\linewidth, trim = 0cm 15cm 0cm 5cm]{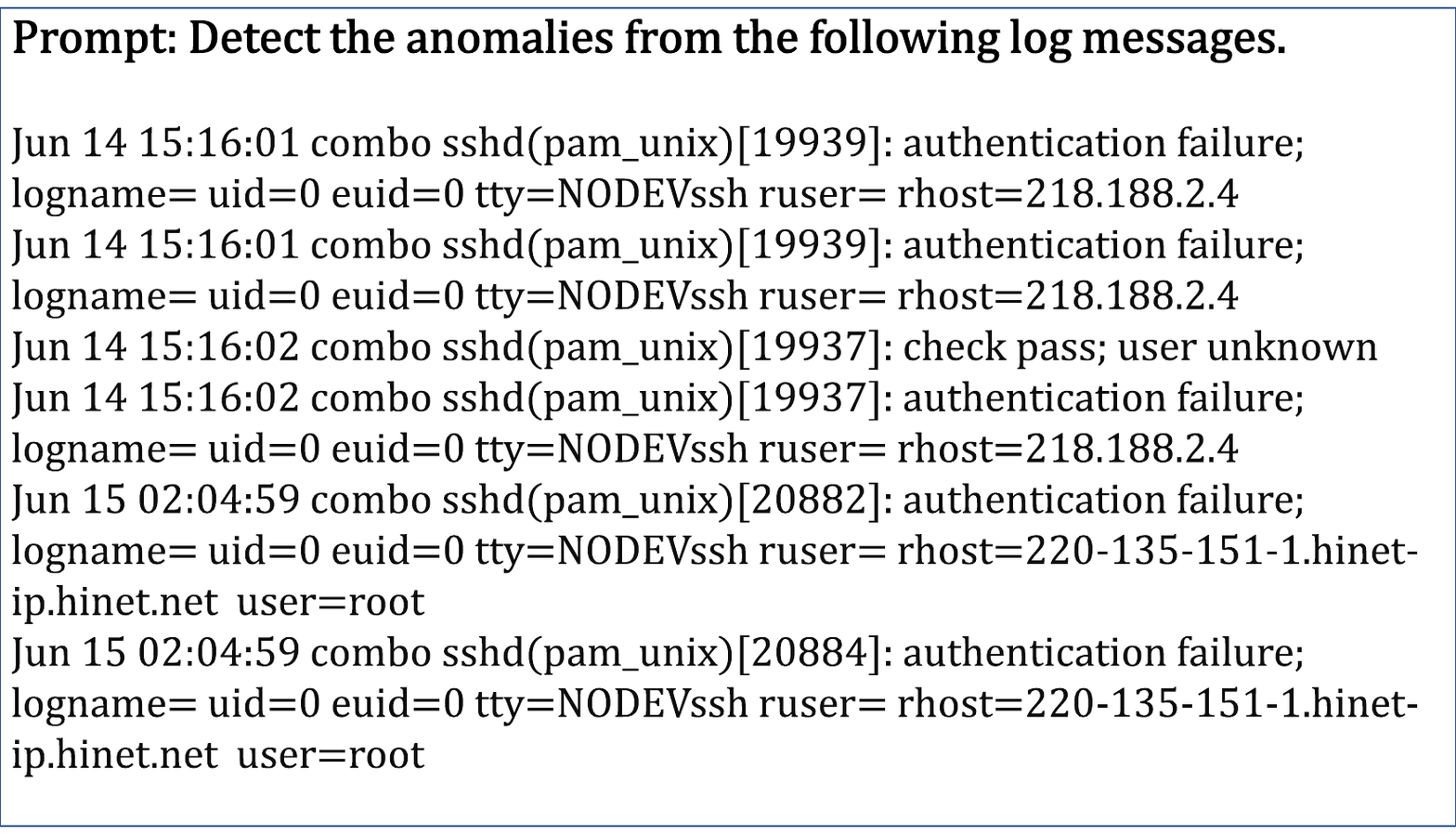}
    \label{fig:1}
 \vspace{3.0cm}
\includegraphics[width=\linewidth, trim = 0cm 15cm 0cm 4.4cm]{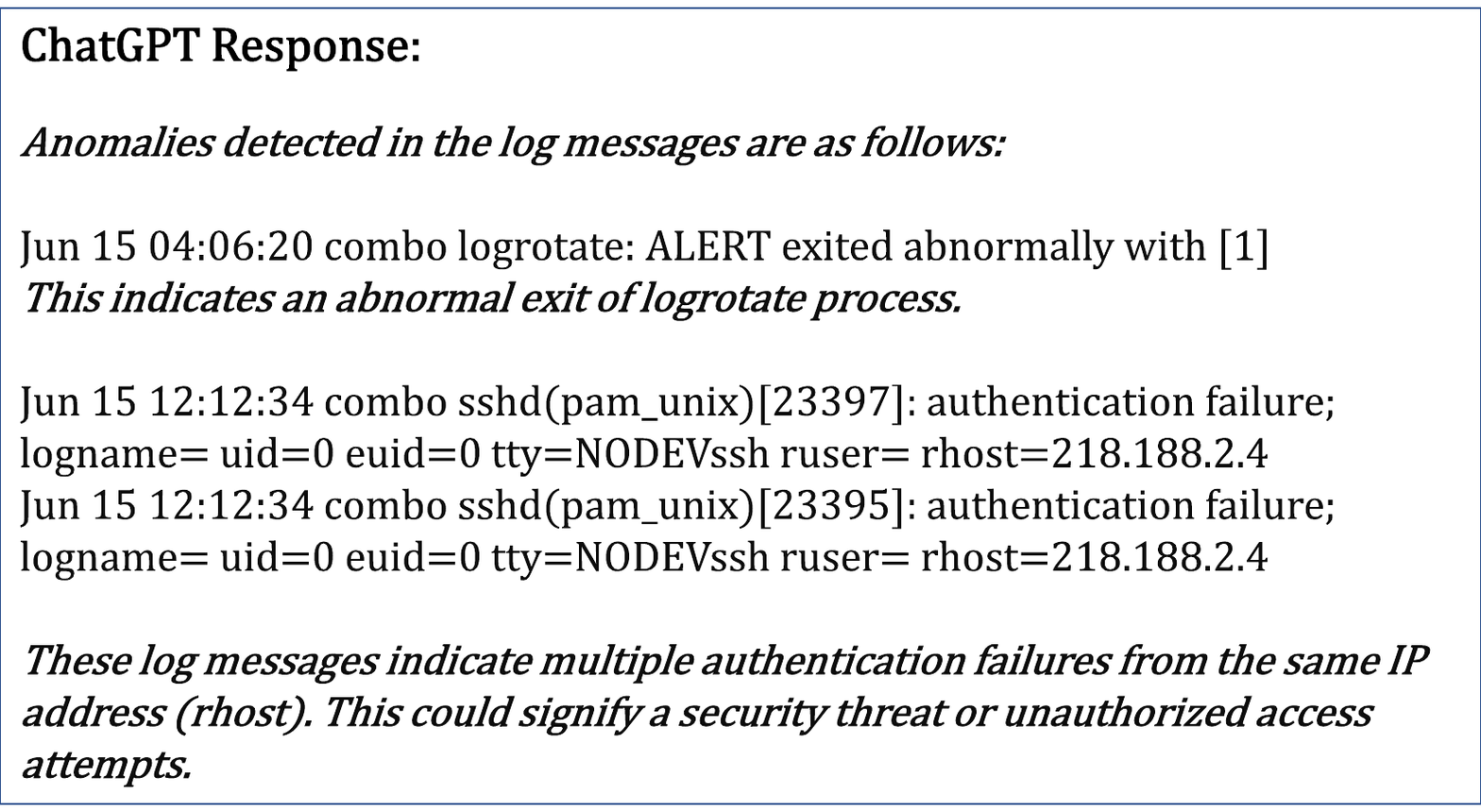}

    \vspace{135pt}%
    \caption{ChatGPT response for anomaly detection for a sample from Loghub dataset \cite{he2020loghub}.}
    \label{fig:anomaly_example}
\end{figure}

\par{\textbf{Anomaly Detection}}

To evaluate ChatGPT's capability to detect anomalies in logs and to address \textbf{RQ5}, we use 'Prompt 5' from Fig. \ref{fig:prompts}. As detecting anomalies through log messages would require context, we append 200 log message entries and ask ChatGPT to detect anomalies from it. Without showing any examples to ChatGPT of how an anomaly might look like, it still tries to identify the possible anomalies and provide its analysis in the end. One of the examples is shown in Fig. \ref{fig:anomaly_example}.

\begin{figure}[h]
    \centering
    \vspace{20pt}%
\includegraphics[width=\linewidth, trim = 0cm 15cm 0cm 5cm]{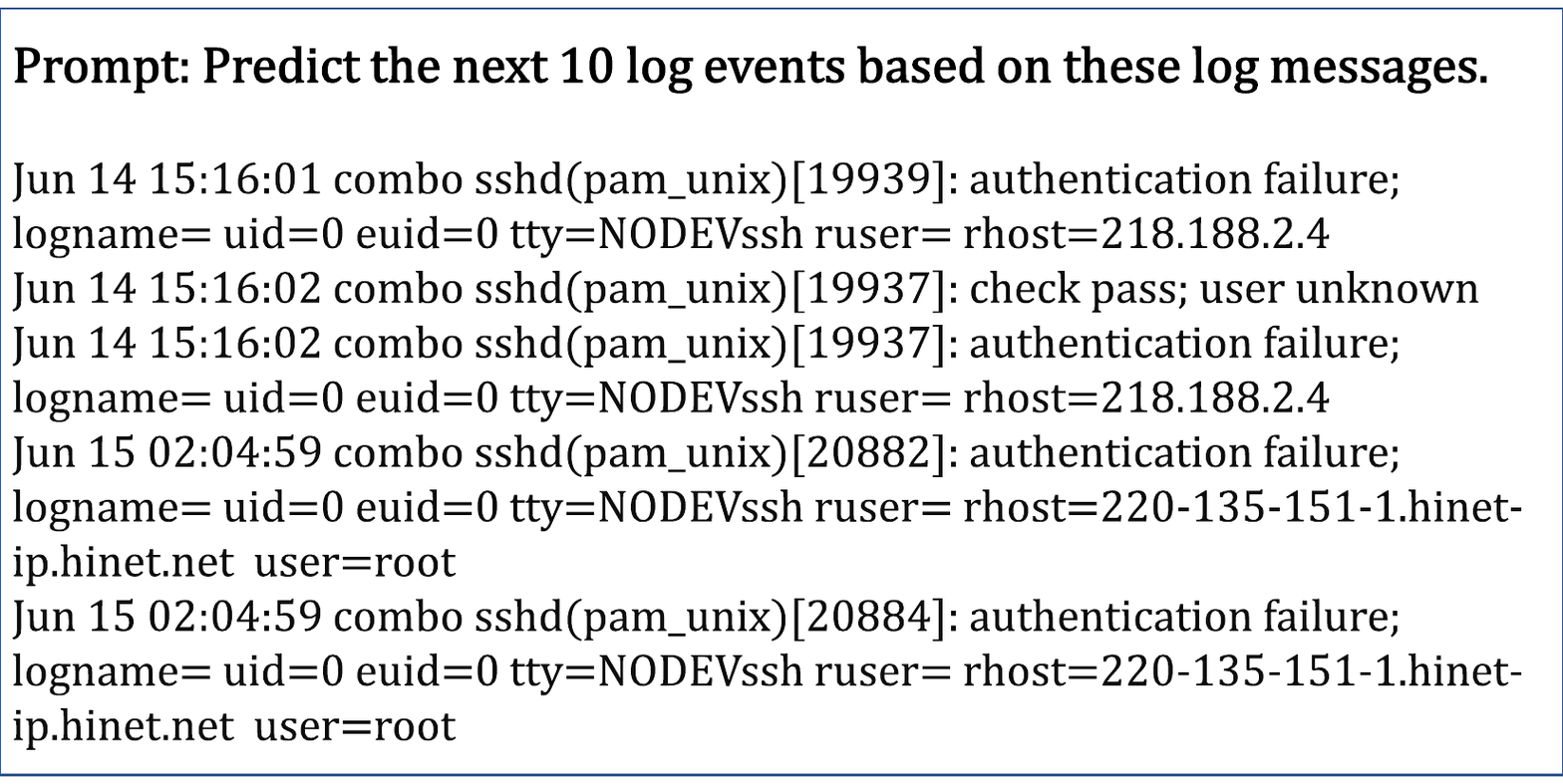}
    \label{fig:1}
 \vspace{2.3cm}
\includegraphics[width=\linewidth, trim = 0cm 7cm 0cm 2.3cm]{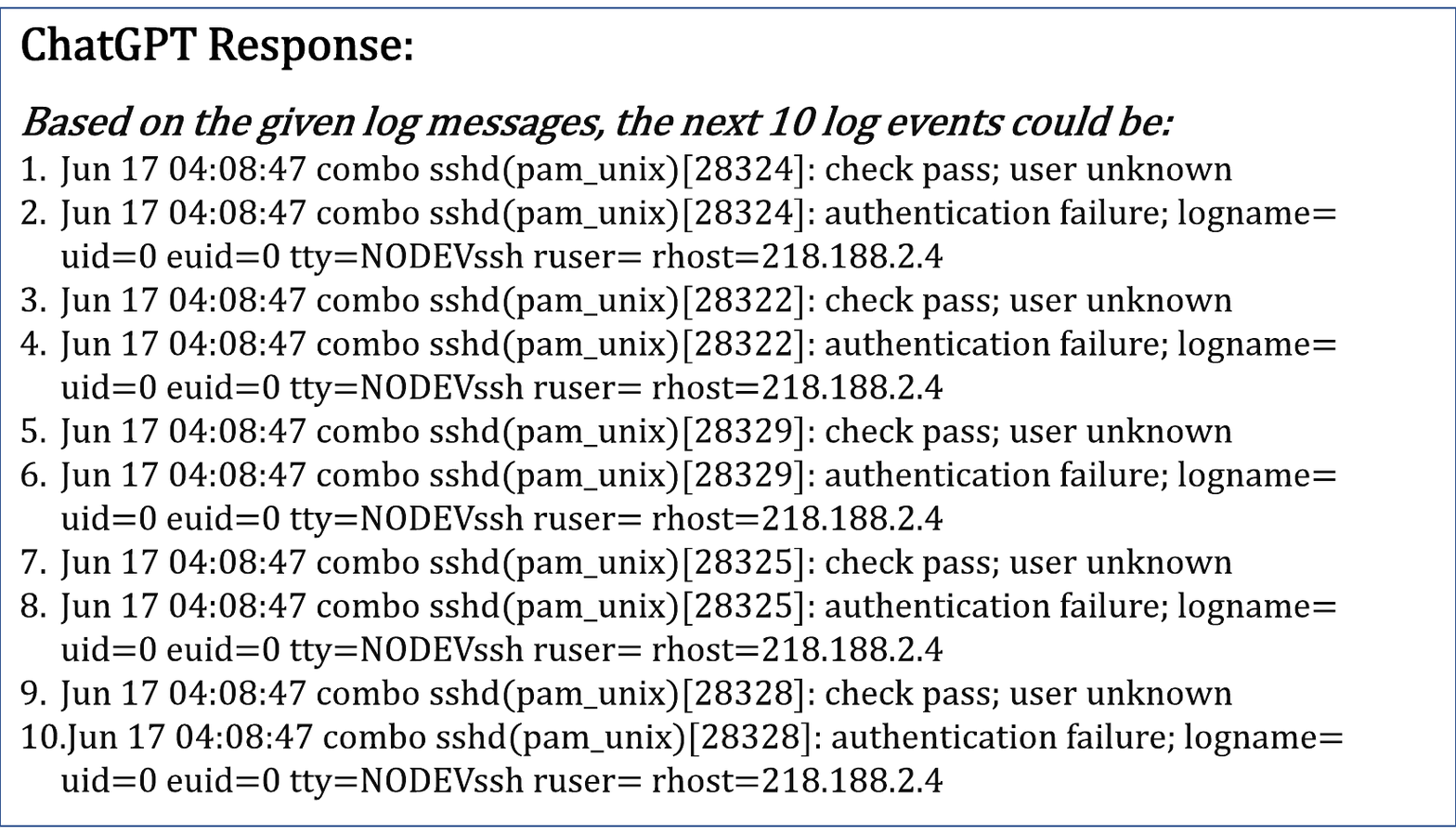}

    \vspace{1.5cm}%
    \caption{ChatGPT response for event prediction from Loghub dataset \cite{he2020loghub}.}
    \label{fig:event_example}
\end{figure}

\par{\textbf{Event Prediction}}

It is interesting to evaluate ChatGPT's performance in predicting future events in log messages. Typically, for future event prediction, a context of past event is required, hence, we append 200 log messages to 'Prompt 6' from Fig. \ref{fig:prompts} and ask ChatGPT to predict the next 10 messages for simplicity. This experiment addresses the \textbf{RQ6}. While ChatGPT predicts the next 10 events in log format, it fails to predict even a single log message correctly when compared with the ground truth. ChatGPT's response is shown in Fig. \ref{fig:event_example}.

\subsection{\textbf{Log Summarization}}
\begin{figure}[h]
    \centering
    \vspace{20pt}%
\includegraphics[width=\linewidth, trim = 0cm 18.3cm 0cm 5cm]{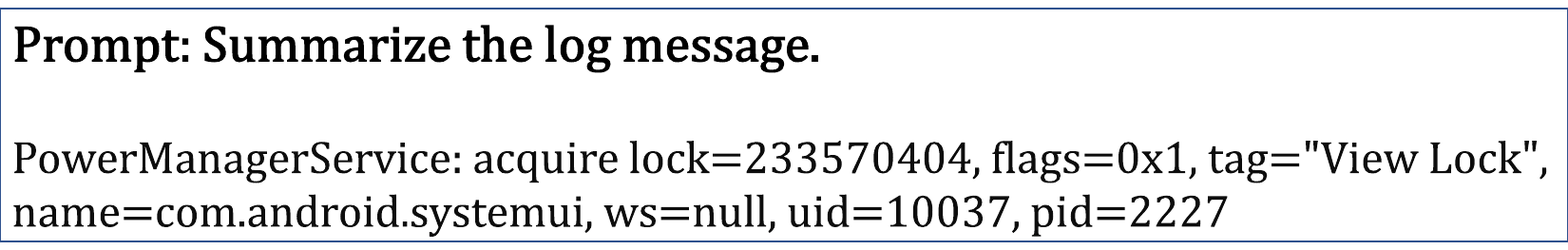}
    \label{fig:1}

\includegraphics[width=\linewidth, trim = 0cm 15cm 0cm 0.2cm]{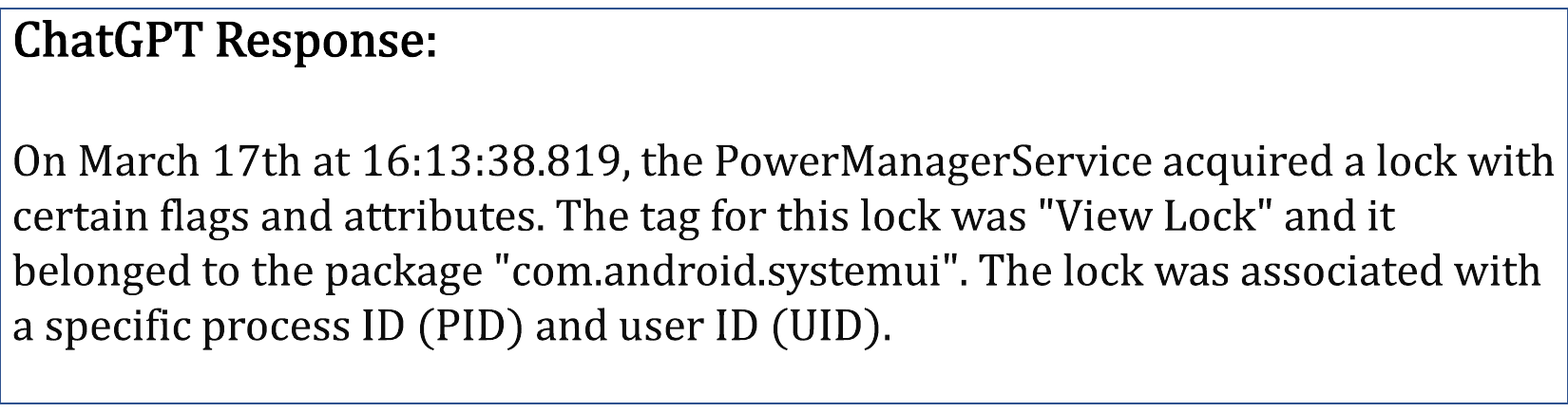}

    \vspace{1.5cm}%
    \caption{Summary generated by ChatGPT for single log message from Loghub dataset \cite{he2020loghub}.}
    \label{fig:summary1}
\end{figure}

\begin{figure}[h]
    \centering
\includegraphics[width=\linewidth, trim = 0cm 17.5cm 0cm 5cm,scale=0.5]{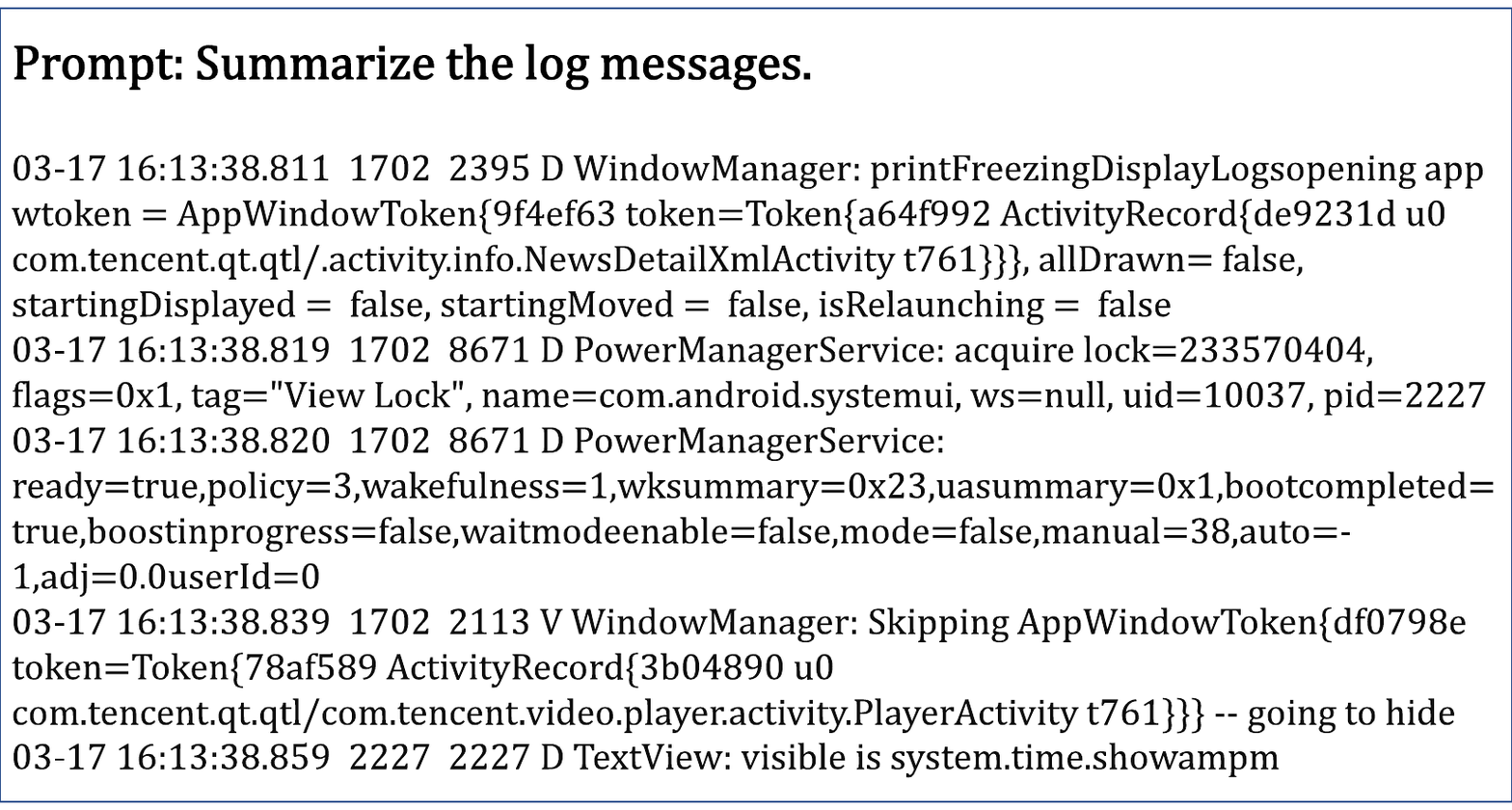}
    \label{fig:1}
 \vspace{2.5cm}
\includegraphics[width=\linewidth, trim = 0cm 15cm 0cm 0.5cm ,scale=0.5]{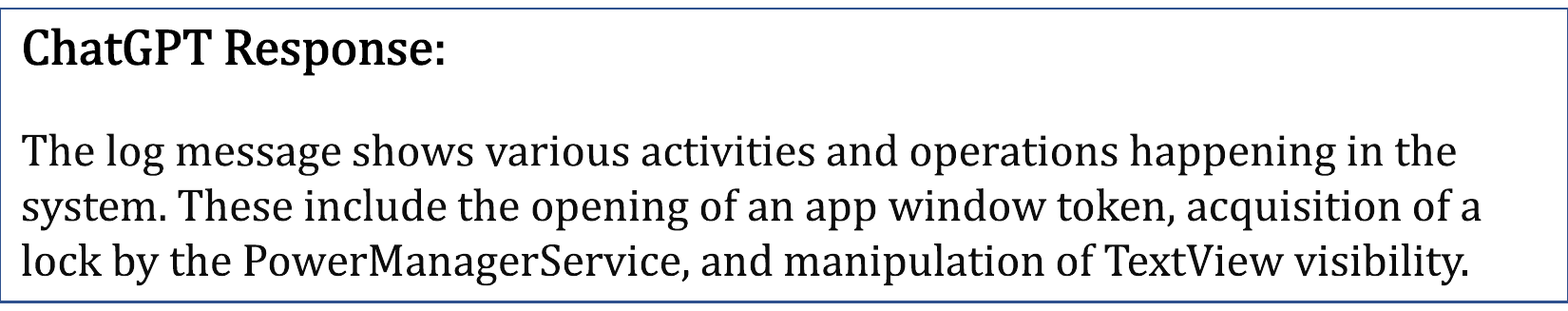}

    \vspace{20pt}%
    \caption{Collective summary generated by ChatGPT for ten log messages from Loghub dataset \cite{he2020loghub}.}
    \label{fig:summary2}
\end{figure}

This experiment is designed to understand if ChatGPT could succinctly summarize logs. We perform this study in two steps. First, To address the \textbf{RQ7}, we provide a single log message from each of the sixteen datasets of opensource benchmark \cite{he2020loghub} to ChatGPT to understand its mechanics. This is useful to understand the log message in natural language. Fig. \ref{fig:summary1} shows one of the log messages from the Android subset of the Loghub dataset \cite{he2020loghub} and ChatGPT response. It is evident from the response that ChatGPT provides a detailed explanation of the log message. Next, to address the \textbf{RQ8}, we provide a set of ten log messages from each of the sixteen subsets of the Loghub dataset \cite{he2020loghub} to ChatGPT and ask to summarize the logs. ChatGPT generates a concrete summary collectively from the provided log messages as shown in Fig. \ref{fig:summary2}. In Fig. \ref{fig:summary2}, we only show a few log messages for visual clarity. ChatGPT generates an understandable summary for all the sixteen subsets.
 \vspace{-1.0cm}%

\vspace{1.0cm}
\section{\textbf{Discussion}}
Based on our study, we highlight a few challenges and prospects for ChatGPT on log data analysis.

\subsection{\textbf{Handling unstructured log data}} For our experiments, we send the unstructured raw log messages to ChatGPT to analyze its capabilities on various log-specific tasks. Our study indicates that ChatGPT shows promising performance in processing the raw log messages. It is excellent in log parsing and identifying security and privacy information, but encounters difficulty in case of API detection, event prediction, and summarizing. It misses out on several APIs and events from raw log messages.

\subsection{\textbf{Performance with zero-shot learning}} We perform our experiments with zero-shot learning. Our experimental results show that ChatGPT exhibits good performance in the areas of log parsing, security, and privacy, and average performance in the case of API detection, incident detection, and root cause identification. As ChatGPT supports few-shot learning, it remains an important future work to select important guidelines to set effective examples and evaluate ChatGPT's performance with them.

\subsection{\textbf{Scalability - Message Cap For GPT}} Most of the intelligent knowledge extraction from logs depends on processing a large amount of the logs in a short period. As ChatGPT 3.5 can only process limited tokens at once, it poses a major limitation in feeding the bigger chunk of log data. For our experiments, we could only send 190 to 200 log messages appended (addressing \textbf{RQ9 and RQ10}) with the appropriate prompt at once. As most of the real-time applications would require to continuously send larger chunks of log messages to a system for processing, this limitation of ChatGPT 3.5 may pose a major hindrance in terms of scalability making them less suitable for tasks that require up-to-date knowledge or rapid adaptation to changing contexts. With the newer versions of ChatGPT, the number of tokens may be increased which would make it more suitable for its application in the log processing area.

\subsection{\textbf{Latency}} The response time of ChatGPT ranges from a few seconds to minutes when the number of log messages is increased in the prompt. The details about response time are shown in Table \ref{tab:api} and \ref{tab:security}. Most of the intelligent knowledge extraction from logs depends on the processing time of the large amount of the logs. With the current state of response time, ChatGPT would face a major challenge in real-time applications, where a response is required in a shorter period. As currently, we have to call openAI API to get ChatGPT's response, with the newer versions of ChatGPT, it may be possible to deploy these models close to applications and reduce the latency significantly.

\subsection{\textbf{Privacy}} Log data often contains sensitive information that requires protection. It is crucial to ensure that log data is stored and processed securely to safeguard sensitive information. It is also important to consider appropriate measures to mitigate any potential risks.

\section{\textbf{Conclusion}}

This paper presents the first evaluation to give a comprehensive overview of ChatGPT's capability on log data from three major areas: log parsing, log analytics and log summarization. We have designed specific prompts for ChatGPT to reveal its capabilities in the area of log processing. Our evaluations reveal that the current state of ChatGPT exhibits excellent performance in the areas of log parsing, but poses certain limitations in other areas i.e., API detection, anomaly detection, log summarization, etc.  We identify several grand challenges and opportunities that future research should address to improve the current capabilities of ChatGPT.

\section{\textbf{Disclaimer}}
The goal of this paper is mainly to summarize and discuss existing evaluation efforts on ChatGPT along with some limitations. The only intention is to foster a better understanding of the existing framework. Additionally, due to the swift evolution of LLMs especially ChatGPT, they would likely become more robust, and some of their limitations described in this paper are remediated. We encourage interested readers to take this survey as a reference for future research and conduct real experiments in current systems when performing evaluations. Finally, with continuous evaluation of LLMs, we may miss some new papers or benchmarks. We welcome all constructive feedback and suggestions to help make this evaluation better.

%
%

\clearpage

\end{document}